\documentclass[12pt]{article}
\usepackage{epsfig}
\usepackage{amsmath}
\usepackage[psamsfonts]{amssymb}
\usepackage{amsthm}
\usepackage{euscript}

\usepackage{latexsym}

\renewcommand{\theequation}{\arabic{section}.\arabic{equation}}
\renewcommand{\(}{\begin{equation}}
\renewcommand{\)}{end{equation} \vspace{-.05in}\linebreak}

\newcounter{saveeqn}
\newcounter{savealpheqn}

\newcommand{\alpheqn}{\setcounter{saveeqn}{\value{equation}}%
 \stepcounter{saveeqn}\setcounter{equation}{0}%
 \renewcommand{\theequation}{\mbox{\arabic{section}.\arabic{saveeqn}\alph{equation}}}
 \renewcommand{\)}{\end{equation}}}
\def\part#1{\frac{\partial}{\partial{#1}}}%
\def\group#1{\refstepcounter{equation}\setcounter{saveeqn}{\value{equation}}%
 \label{#1}\setcounter{equation}{0}%
 \renewcommand{\theequation}{\mbox{\arabic{section}.\arabic{saveeqn}\alph{equation}}}
 \renewcommand{\)}{\end{equation}}}
\newcommand{\reseteqn}{\setcounter{equation}{\value{saveeqn}}%
 \renewcommand{\theequation}{\arabic{section}.\arabic{equation}}%
 \renewcommand{\)}{\end{equation}}}
\newcounter{alphcount}

\def\writeletter#1{\renewcommand{\theequation}{\alph{#1}}%
                 \begin{eqnarray}%
                 \label{#1}%
                 \nonumber\end{eqnarray}\vspace{-.666in}}
\def\getlette2r#1{\newcounter{#1}%
                 \setcounter{#1}{\value{equation}}%
                 \providecommand{\writeletters}{\writeletters\writeletter{#1}}}

\newcommand{\aalpheqn}{\setcounter{saveeqn}{\value{equation}}%
 \stepcounter{saveeqn}\setcounter{equation}{0}%
 \renewcommand{\theequation}{\mbox{\Alph{subsection}.\arabic{saveeqn}\alph{equation}}}
  \renewcommand{\)}{\end{equation}}}
\newcommand{\areseteqn}{\setcounter{equation}{\value{saveeqn}}%
 \renewcommand{\theequation}{\Alph{subsection}.\arabic{equation}}%
 \renewcommand{\)}{\end{equation}}}

\renewcommand{\thefootnote}{\alph{footnote}}
\renewcommand{\(}{\begin{equation}}
\renewcommand{\)}{\end{equation}}
\newcommand{\ba}{\begin{eqnarray}}
\newcommand{\ea}{\end{eqnarray}}

\newcommand{\bp}{\mathop{\vtop{\ialign{##\crcr
  $\hfil\displaystyle{}\hfil$\crcr\noalign{\kern-13pt\nointerlineskip}
  \BIG{(}\hskip0pt\crcr\noalign{\kern3pt}}}}}
\newcommand{\cbp}{\mathop{\vtop{\ialign{##\crcr
  $\hfil\displaystyle{}\hfil$\crcr\noalign{\kern-13pt\nointerlineskip}
  \BIG{)}\hskip0pt\crcr\noalign{\kern3pt}}}}}
\newcommand{\pa}{\mathop{\vtop{\ialign{##\crcr
  $\hfil\displaystyle{\oplus}\hfil$\crcr\noalign{\kern+1pt\nointerlineskip}
  \hspace{.08in}$^{\alpha=0}$\hskip6pt\crcr\noalign{\kern3pt}}}}}
\renewcommand{\sp}{,\hspace{.3in}}
\newcommand{\newsection}{\setcounter{equation}{0}\section}
\newcommand{\p}{^\prime}

\newcommand{\R}{\ensuremath{\mathbb R}}

\newcommand{\Z}{\ensuremath{\mathbb Z}}

\newcommand{\del}{\ensuremath{\partial}}

\newcommand{\beq}{\begin{equation}}
\newcommand{\eeq}{\end{equation}}

\def\BIG#1{\mbox{\Huge $#1$}}

\mathchardef\endbar="375
\font\fivesans=cmss10 at 4.61pt
\font\sevensans=cmss10 at 6.81pt
\font\tensans=cmss10 at 12pt 
\newfam\sansfam
\textfont\sansfam=\tensans\scriptfont\sansfam=\sevensans\scriptscriptfont
\sansfam=\fivesans
\def\sans{\fam\sansfam\tensans}
\def\Z{{\mathchoice
{\hbox{$\sans\textstyle Z\kern-0.455em Z$}} 
{\hbox{$\sans\textstyle Z\kern-0.455em Z$}} 
{\hbox{$\sans\scriptstyle Z\kern-0.355em Z$}} 
{\hbox{$\sans\scriptscriptstyle Z\kern-0.255em Z$}}}} 

\font\tensans=cmss10 at 14pt 
\newfam\sansfamb
\textfont\sansfamb=\tensans\scriptfont\sansfamb=\sevensans\scriptscriptfont
\sansfamb=\fivesans

\font\tensans=cmss10 at 17pt
\newfam\sansfamc
\textfont\sansfamc=\tensans\scriptfont\sansfamc=\sevensans\scriptscriptfont
\sansfamc=\fivesans

\def\contr#1#2{\mathop{\vtop{\ialign{##\crcr
  $\hfil\displaystyle{#2}\hfil$\crcr\noalign{\kern3pt\nointerlineskip}
  \hspace{.09in}\rule[0in]{.01in}{.1in}\rule[0in]{#1in}{.01in}\rule[0in]{.01in}{.1in}\hskip6pt\crcr\noalign{\kern3pt}}}}}
\def\contrb#1#2#3{\mathop{\vtop{\ialign{##\crcr
  $\hfil\displaystyle{#3}\hfil$\crcr\noalign{\kern3pt\nointerlineskip}
  \hspace{#1in}\rule[0in]{.01in}{.1in}\rule[0in]{#2in}{.01in}\rule[0in]{.01in}{.1in}\hskip6pt\crcr\noalign{\kern3pt}}}}}

\def\hsp#1{\hspace{#1in}}

\catcode`\@=11
\def\vereq#1#2{\lower3pt\vbox{\baselineskip1.5pt \lineskip1.5pt
\ialign{$\m@th#1\hfill##\hfil$\crcr#2\crcr\sim\crcr}}}
\catcode`\@=12

\newtheorem{conjecture}{Conjecture}

\begin{document}
\begin{titlepage}
\begin{center}
Friday the 13th of October, 2000        \hfill UCB-PTH-00/35  \\
v2: Good Friday the 13th of April, 2001 \hfill hep-th/0010092 \\


\vskip 1in
\def\thefootnote{\fnsymbol{footnote}}

{\large \bf Backreaction I: The Torus\\}

\vskip 0.3in

Jarah Evslin\footnote{E-Mail: jarah@uclink4.berkeley.edu}, Uday Varadarajan\footnote{E-Mail: udayv@socrates.berkeley.edu} and John E. Wang\footnote{E-Mail:
  hllywd2@physics.berkeley.edu}

\vskip 0.15in

{\em Department of Physics,
     University of California\\
     Berkeley, California 94720}\\
       
\end{center}


\vfill

\begin{abstract}
  
  We use wrapped D-brane probes to measure position dependent
  perturbations of compactification moduli. Due to the backreaction of
  the D-branes on the local geometry, we suspect that measuring the
  fluctuations of one modulus to high precision will generically affect
  the others.  These considerations lead us to conjecture a novel
  uncertainty principle on the Calabi-Yau moduli space.  We begin our
  investigation with a gedanken experiment on a torus.

\end{abstract}

\vfill

\end{titlepage}
\setcounter{footnote}{0}
\renewcommand{\thefootnote}{\arabic{footnote}}

\pagebreak
\renewcommand{\thepage}{\arabic{page}}
\pagebreak 
\newsection{Motivation} 

Among the most fascinating and longstanding puzzles that physicists
face in this new century is: ``What is the nature of spacetime
geometry?''  The classical, (pseudo-)Riemannian geometry that served
us during the last century does not adequately describe the universe
as seen by strings and D-branes.  Classical geometry fails to capture
the smooth behavior of conformal field theories at orbifold
singularities and flop transitions \cite{WittenFlop, AspinwallFlop},
as well as the smooth physics of conifold singularities that are
resolved \cite{Strominger} or subject to discrete torsion \cite{Vafa,
  Douglas} in the presence of D-branes.  Every shortcoming of
classical geometry is a clue to this puzzle, and as these pieces are
assembled, a picture of a new stringy geometry \cite{WittenFlop,
  AspinwallFlop, Douglas, GreeneSTCY} is only beginning to emerge.

Consider a spacetime consisting of a noncompact $D$-dimensional
manifold, $M$, with a compact Calabi-Yau manifold over each point.
Classically, each simply-connected, compact Calabi-Yau manifold is
described by some topological characteristics (Hodge numbers, an
intersection form, etc.) and also a point in moduli space, which
corresponds to a choice of K\"{a}hler and complex structure moduli.
In stringy geometry there are smooth transitions which connect the
moduli spaces of topologically distinct Calabi-Yau, forming an
\textit{extended moduli space}.  Thus our spacetime is described by
associating a point in the extended moduli space with each point in
$M$.  Even in the simplest case, where the moduli are chosen
identically everywhere, many new features of stringy geometry have
been discovered.

However some phenomena only appear when the moduli are allowed to
vary. For example, near large numbers of BPS D-branes wrapped around
non-trivial cycles in a Calabi-Yau, we know that the moduli experience
an attractor flow \cite{FerraraAttract, FerraraAttract2,
FerraraAttract3} as one moves toward the branes. To better understand
the nature of stringy geometry, we will consider such variations in
moduli as seen by D-brane probes. In particular, we will pose the
question, ``How well can D-branes measure local variations in
moduli?''

We will present arguments to motivate the following conjecture.
\begin{conjecture}
A measurement of a variation in a period\footnote{The periods $p_i$
are a set of coordinates on moduli space that roughly correspond to
the volume of cycles $C_i$ in the Calabi-Yau.} $p$ which is localized
to a region of diameter $d$ in the noncompact directions will
generically affect the values of the other periods.  Furthermore this
effect increases as the precision of the measurement of $p$ increases
and as $d$ decreases.
\end{conjecture}
In particular we expect that for $d$ very large, corresponding to
measuring a nearly constant modulus, the conjectured effect
disappears.  Assuming that this phenomenon is not an artifact of our
measurement scheme, we can interpret it as an uncertainty principle.

\begin{figure}[htb]
\begin{center}
\epsfxsize=6in\leavevmode\epsfbox{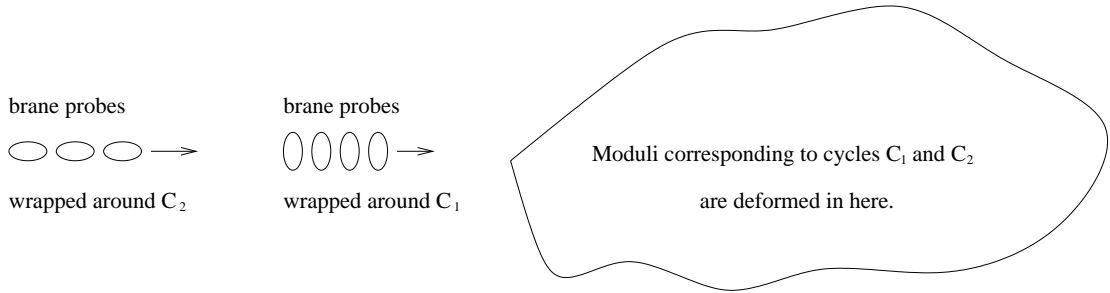}
\caption{Branes measure period $p_1$ then $p_2$.  Do these operations
  commute?} \label{FigProbes}
\end{center}
\end{figure}
 
This conjecture is plausible for the following reason.  Imagine that
we measure a local variation of period $p_1$ by wrapping branes
around the corresponding cycle $C_1$ and using these branes to probe
the region with the variation (see Fig.~\ref{FigProbes}).  To obtain an
accurate measurement of the deformation of $p_1$, we slowly
send many branes very close to the region to be measured.
However this is likely to cause a backreaction on the geometry, which
(using attractor flows as a guide) we suspect will alter the other 
periods in much the same way as an
accurate position measurement alters momentum in the case of the
Heisenberg uncertainty principle.

Being mere mortals, in this paper we will only begin an analysis of
the validity of this conjecture.  We will consider an illustration of
our proposed experiment in the case of toroidal compactification.  In
the case of the two-torus, we find a variant of the space-space
uncertainty expected in any theory of quantum gravity, which upon
dimensional reduction becomes a Heisenberg uncertainty principle.  We
conclude with preliminary evidence for our conjecture on the
four-torus.

The next step in this analysis would be to apply the
lessons learned below to the measurement of moduli on $T^4$,
$T^6$, $T^2\times K3$, as well as some simple orbifolds and
multi-parameter Calabi-Yau three-folds if possible.

\newsection{The Experiment}

We consider Type II string theory at small $g_s$ and in a spacetime of
the form $\R^{1,D-1} \times T^2$.  A critical string theory may be
obtained by adding a superconformal field theory with central charge $c=12-(3D/2)$.  We perturb
spacetime so that the two radii of the torus vary in the noncompact
directions, which yields a torus fibration rather than a direct
product. Specifically, the tori have constant radii
\begin{equation}
R_1=R_2=R \gg\sqrt{\alpha\p}
\end{equation}
everywhere in $\R^{1,D-1}$ except for a small,
spherical region in space of diameter $d$. In the interior of this
sphere the radii of the torus are slightly different from $R$, each by
a small amount (not necessarily the same amount for both radii) of
order $\delta R \ll R$. We describe this variation perturbatively as a
wavepacket consisting of a superposition of closed strings.

The moduli describing these variations are massless fields in the low
energy effective theory on $\R^{1,D-1}$ so this wavepacket travels at
the speed of light and is composed of strings with momenta on the
order of $1/d$.  As these strings interact with each other, the
wavepacket will tend to disperse. This dispersion is expected, as the
wavepacket that we are considering is a finite energy perturbation of
the interacting theory which does not, a priori, correspond to a
stable deformation of the string background\footnote{However, in
Ref.~\cite{Rudd} it was shown that smooth variations of the moduli of
the torus which admit a timelike killing isometry can be lifted to
exact classical string vacua by adding an appropriate compensating
variation of the dilaton and light-cone gauge metric, at least if
$D=2$.}.


There are perturbative arguments that lead us to believe that these
variations in moduli are sufficiently stable for the experiment to be
sensible. To see this, note that the leading contribution to this
dispersion comes from the four point function on the sphere, which is
of order $g_s^2$ and not from the on-shell three point amplitude,
which is kinematically forbidden.  One advantage of using D-brane
probes is that they interact with the moduli with an amplitude of
order $g_s$, and therefore to leading order in $g_s$, we may neglect
the dispersion. In addition, for small $\delta R$, the density of closed
strings in the packet is small.  This suppresses the dispersion
further because the dispersion scales as the density squared, while
our measurement process scales only as the density.

To measure the variation $\delta R$, let us place an advanced
experimental physicist in the path of the wavepacket.  He knows both
the packet's trajectory and the order of magnitude of its size, but
wants to determine $\delta R$ more precisely.  The experimenter wants
to understand spacetime geometry as the branes see it, and so he is
only interested in the scattering of the wavepacket by branes and not
by any other probes at his disposal.  Fortunately he has a reservoir
of branes of various dimensions and wrappings in his lab which he can
arrange in any configuration before the wavepacket arrives.  After
reading Refs.~\cite{FriedanCFT, KlebanovNS, KlebanovRR, Garousi,
  KlebanovReview}, he figures out how many branes of each type he will
need so that on the order of one brane will be scattered by the
incoming wavepacket.

\begin{figure}[htb]
\begin{center}
  \epsfxsize=6in\leavevmode\epsfbox{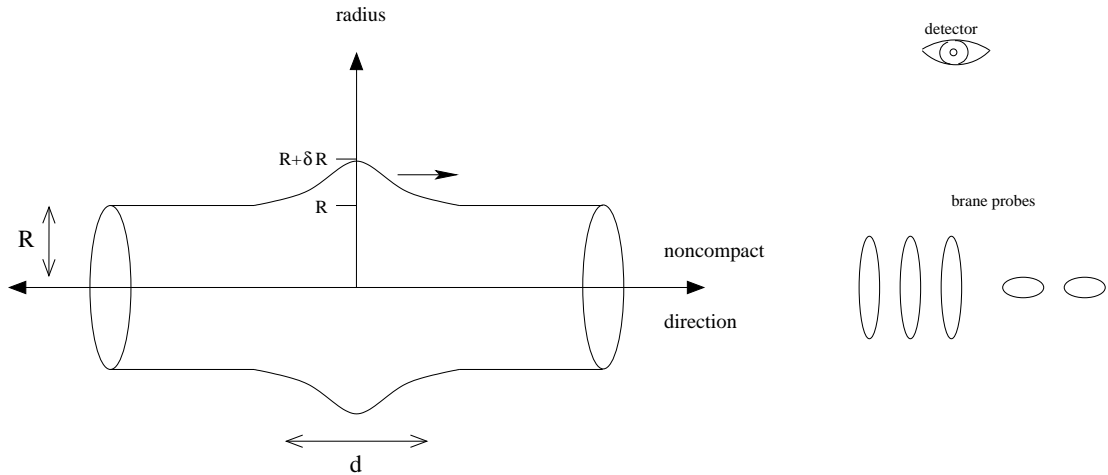}
\caption{A variation knocks brane probes into a detector.} \label{FigVariation}
\end{center}
\end{figure}

The branes are placed in a line, along the path of the variation (see
Fig.~\ref{FigVariation}). The tree-level scattering calculations
\cite{Garousi, KlebanovReview} are only valid when spacetime is
reasonably flat, which is achieved by separating the brane probes
sufficiently.  Eventually the packet passes through the assortment of
branes, knocking some of them into a detector. After counting hits and
reconstructing the time of the scattering from the brane velocities,
our experimenter extrapolates $\delta R$.  However in order to obtain
a very accurate measurement, the experimenter would need a large
number of branes, and this would deform the modulus variation that he
is trying to measure so much that his results would be useless.  For
example, the first few branes would see a very different wavepacket
from the last few.

That is to say, there is an uncertainty bound characteristic of
theories of quantum gravity (not the uncertainty we have conjectured)
on how accurately he can perform this measurement. For the duration of
this paper we will describe just what the branes do to these moduli
and what kind of uncertainty relation this implies.

\newsection{The Calculation} \label{SecCalc}

To analyze the preceding gedanken experiment we have calculated the
relevant scattering amplitudes using the results of Refs.~\cite{Garousi,
  KlebanovReview}.  Specifically, for the case of flat space and thus
toroidal compactifications, the authors calculated tree-level
amplitudes for closed strings scattering off D-branes.  We will
outline those calculations here with a view towards the analogous
calculations in non-trivial Calabi-Yau string compactifications.

\subsection{The Scattering}

\begin{figure}[htb]
\begin{center}
  \epsfxsize=6in\leavevmode\epsfbox{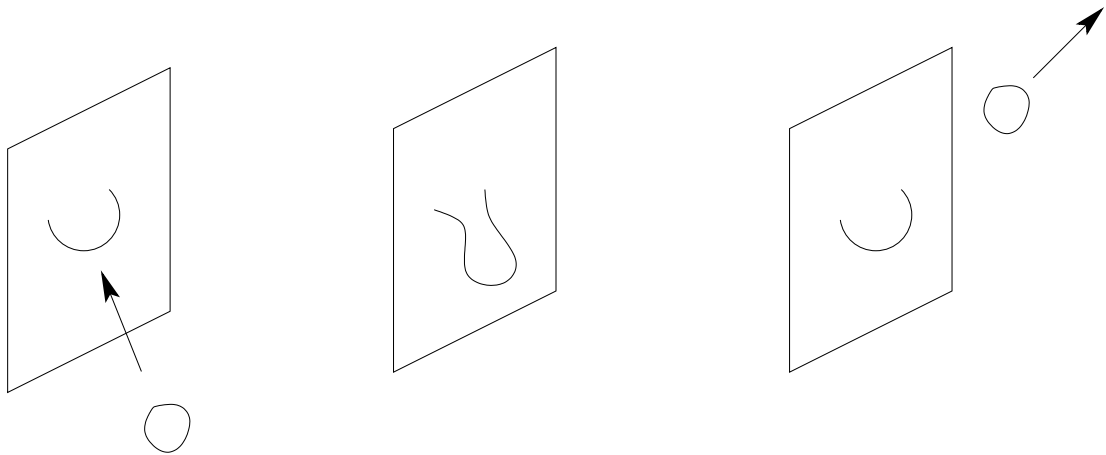}
\caption{Closed strings scatter off a 2-brane.} \label{FigClosedString}
\end{center}
\end{figure}

We are interested in the interaction of a closed string comprising
the modulus and a D-brane completely wrapped around some toroidally
compactified dimensions.  The relevant tree-level calculation is the
absorption and re-emission of a closed string from an open string
attached to the brane seen schematically in Fig.~\ref{FigClosedString}.
Thus we are interested in a disk with two punctures, corresponding to
the vertex operators of the incoming and outgoing closed strings.  The
incoming modulus is a massless NS-NS closed string with polarizations
in the compact directions and momentum only in the noncompact
direction.  The outgoing state, at lowest order in $g_s$, may, a
priori, be any massless (in the $D+2$-dimensional sense) closed string
with arbitrary polarization and momenta in either the compact or
noncompact directions (although we will see that, as expected, some
final states are forbidden) or a pair of open strings attached to the
brane.

As an example, consider the case in which both the initial and final
states are NS-NS closed strings. Their vertex operators are of the
form
\begin{equation}
V= \epsilon_{\mu \nu} (\partial X^{\mu} + ip \cdot \psi \psi^{\mu})
(\bar{\partial} X^{\nu} + ip \cdot \bar{\psi} \bar{\psi}^{\nu}) e^{ip
  \cdot X} 
\end{equation}
where $p$ is the momentum and $\epsilon$ a polarization tensor.  

We always assume that the incoming string has no compact momentum.
The compact momentum of the outgoing string is quantized, $p_m=N_m/R$,
where the index $m$ runs over the compact coordinates. To
specify the vertex operators for the incoming moduli, we only need to
identify their polarization tensors. The K\"{a}hler modulus is a
complex combination of the volume of the torus and the integral of the
NS-NS two form over the torus, while the complex structure modulus can
be identified by writing the metric on the torus in standard form as
\begin{equation}
ds^2 = \left| dx_{D} + \tau dx_{D+1} \right|^2.
\end{equation}
Thus, we can arrive at the appropriate form for the polarization
tensors by writing small deformations of these moduli in terms of
variations of the metric, dilaton and antisymmetric tensor fields.
We then find that polarization tensors corresponding to incoming
excitations of K\"{a}hler and complex moduli can be written as
\begin{equation}
\epsilon^K= \left( \begin{array}{ll} \hsp{.1} i & 1 \\ -1 & i \end{array}
\right)\sp
\epsilon^C= \left( \begin{array}{ll}  i & \hsp{.1} 1 \\ 1 & -i \end{array} \right) \label{vops}
\end{equation}
in the compact directions and zero in the noncompact directions.

Since toroidal compactifications are described by free field theories on the
world-sheet, the vertex operators were easy to construct.  However, in
a general Calabi-Yau compactification the vertex operators we are
interested in correspond to truly marginal deformations of the SCFT
and are more difficult to construct.  Fortunately many Calabi-Yau
moduli spaces have distinguished points where the nonlinear sigma
model is a rational $(2,2)$ superconformal field theory, such as the
Gepner point on the quintic hypersurface.  Near such points we can
construct these deformations from the (chiral, chiral) and (chiral,
antichiral) primaries as described in Ref.~\cite{GreeneSTCY}.
Boosting the resulting vertex operator in a noncompact direction
yields our wavepacket. One can explicitly check that in the case of
toroidal compactification this procedure yields the polarizations
given in Eq.~(\ref{vops}).


Once the appropriate vertex operators have been constructed, the
amplitude for the disk with two closed string insertions (and one
superconformal ghost, as required on the disk) is calculated in
Refs.~\cite{Garousi, KlebanovReview} using the covariant formulation
of Ref.~\cite{FriedanCFT}. By imposing Neumann boundary conditions in
$p+1$ directions and Dirichlet conditions on the rest, the authors
obtain an amplitude for the scattering of an NS-NS string by the
p-brane in flat space. We will not reproduce their formulas here, but
instead refer the interested reader to Refs.~\cite{Garousi,
KlebanovReview} for details.  Although the final state may be in the
R-R sector (or even a pair of open string excitations of the brane
when $d<R$), for our present purpose it will suffice to restrict our
attention to NS-NS outgoing states.


\subsection{Results}

We first consider the case $d > R$, so that the moduli in the
wavepacket have insufficient energy to excite any internal modes of
the brane or Kaluza-Klein modes on the torus.  In this case, we found
that 0 and 2 branes interact with K\"{a}hler moduli, while 1 branes
interact with complex structure moduli.  This is expected since the
K\"{a}hler modulus on a torus is described by the period of a closed
(1,1) form while the complex structure modulus can be described by the
periods of closed (0,1) and (1,0) forms.  Generally, the branes scatter
these moduli either to R-R states or to gravitons and dilatons with
both polarizations along the noncompact directions, but never mix the
K\"{a}hler and complex structure moduli.  When the final polarizations
of the NS-NS states are both along the noncompact directions, one of
the outgoing polarizations must (using the standard $\epsilon\cdot
p=p\cdot\epsilon=0$ gauge) lie along the plane of the scattering and
be orthogonal to the outgoing momentum.  We will use this fact in
Sec.~\ref{t4} when we consider measurements on $T^4$.  This and other
possible scattering processes are summarized in the table below.

\vspace{.3in}
\hspace{-.3in}
\begin{tabular}{c|c|l|l}
Brane Dim.&d vs. R&Incoming Modulus&Possible Outgoing (NS-NS) States\\ \hline
0 or 2&$d>R$&K\"{a}hler&Dilaton, Graviton, K\"{a}hler Modulus\\
0 or 2&$d>R$&Complex Str.&No Interaction\\
1&$d>R$&K\"{a}hler&No Interaction\\
1&$d>R$&Complex Str.&Dilaton, Graviton, Complex Str. Modulus\\
0,1 or 2&$d<R$&K\"{a}hler&Any\\
0,1 or 2&$d<R$&Complex Str.&Any\\ 
\multicolumn{4}{c}{}\\
\multicolumn{4}{c}{Table 1: Possible Scattering Processes} 
\end{tabular}

\vspace{.3in}

When the size of the wavepacket, $d$, is less than the largest radius
of the torus we can excite Kaluza-Klein modes.  In this case we find
that the scattering proceeds as one would expect in the flat space
case, with the branes interacting generically with all massless closed
string states.  Outgoing closed strings with non-vanishing compact
momenta can have any transverse polarization, and at the same order in
$g_s$ the final state can consist of two open strings bound to the
brane probe with opposite compact momenta.

\newsection{Space-Space Uncertainty Principle}

In the previous section we saw that trying to measure $\delta R$
generically transformed the polarizations of the closed strings in our
wavepacket, thus changing the value of the modulus.  Below we argue
that this indicates the existence of a version of the space-space
uncertainty relation present in theories of quantum gravity.  We will
derive our version of this relation in three different ways.  For
simplicity we will restrict our attention to dilations of the area of
the torus, the quanta of which we will refer to as dilatons.

Our experimenter has chosen the number of branes so that of order one
brane will scatter.  This means that, because the wavepacket is
assumed to be dilute, of order one dilaton will scatter.  Even for the
case when the Kaluza-Klein modes are suppressed ($d>R$), the outgoing
string will generically have polarizations along the noncompact
directions.  Thus $\delta R$ is changed by the loss of one incoming
dilaton.  As a result the smallest $\delta R$ that our experimentalist
can measure at fixed $d$, corresponds to a wavepacket containing of
order one dilaton.  To learn what this $\delta R$ is, we will
count the number of dilatons in a packet.

We know that the dilatons have energies of order $1/d$, thus it will
suffice to calculate the energy of a given packet.  The energy of a
packet $E_{\textup{\scriptsize{packet}}}$ should depend on its size $d$
and its amplitude $\delta R$ and so we can solve for the smallest
measurable $\delta R$ as follows:
\begin{equation} 
1\sim\textup{\# of
  dilatons}=\frac{E_{\textup{\scriptsize{packet}}}}{\textup{energy per
    dilaton}}\sim d\times E_{\textup{\scriptsize{packet}}}\hsp{.1}.
\end{equation}
We will now calculate the total energy as a function of $d$ and
$\delta R$ in three ways.
 
\subsection{Counting Dilatons}

We will calculate the energy of the variation using the effective field
theory, Einstein's equation, as well as a model of space as a
stretchable sheet with constant tension equal to the inverse Newton's
constant, $T=1/G_N$.

Consider the low energy effective theory for the NS-NS sector of Type
II string theory compactified on a two-torus. We assume that the
non-compact space is flat $\R^{1,D-1}$ (with $D=8$ in the critical
case).  In this theory, the moduli are realized as combinations of the
various scalar fields arising from the components of the metric and
NS-NS antisymmetric tensor field along the torus. The K\"{a}hler
modulus is a complex scalar field which is a combination of the
determinant of the metric on the torus $\sqrt{\det(g_{mn})}$ and the
antisymmetric tensor field component $B_{mn} = -B_{nm}$ along the
torus , while the complex structure modulus is a combination of the
metric components $g_{mn}$. Explicitly, the K\"{a}hler and complex
structure moduli can be identified as the complex scalars $\rho =
\rho_1 + i\rho_2$ and $\tau = \tau_1 + i \tau_2$ which are defined by
(take D=8 for ease of notation):
\begin{equation} \label{modfields}
\rho = \frac{R^2}{\alpha'}\left(B_{89} + i \sqrt{\det(g_{mn})}\right)\sp
\tau = \frac{1}{g_{88}}\left({g_{89}+i\sqrt{\det({g_{mn}})}}\right).
\end{equation}
Note that the imaginary part of $\rho$ is related to area of the
torus by
\begin{equation}
\rho_2 = \frac{A}{4\pi^2 \alpha\p} \hsp{.1} .
\end{equation}
Thus, to consider the energy associated to a localized perturbation of
the area of the torus moving along the noncompact $x_1$ direction, with some
magnitude associated with a radius perturbation $\delta R$, we can take
\begin{equation} \label{myrho}
\rho_2 = \frac{A_0}{4\pi^2 \alpha\p} \left( 1 + \frac{\delta
        R}{\sqrt{A_0}} \exp \left[-\frac{(x_1-t)^2+|{\bf x}|^2}{d^2}
        \right] \right)
\end{equation}
(where $A_0$ is the area of the unperturbed torus) and calculate the
energy of the configuration in the effective theory.

To do this, we first note that the effective action for the massless
fields in the NS-NS sector in a toroidally compactified type II string
theory, expressed in Einstein frame, takes the form
\begin{equation}
S = \frac{1}{16 \pi G_N^{D}} \int d^{D}x (-G_D)^{1/2}\left( R_D
  -\frac{1}{2} \frac{\del_{\mu}\tau  \del^{\mu}\bar{\tau}}{\tau_2^2}
  -\frac{1}{2} \frac{\del_{\mu}\rho  \del^{\mu}\bar{\rho}}{\rho_2^2} +
  \cdots \right).
\end{equation}
Here $G_N^D$ is the $D$-dimensional Newton's constant, while $G_D$ and
$R_D$ are the effective $D$-dimensional spacetime metric and Ricci
scalar respectively.  To compute the energy of the wavepacket, we make
the field redefinition $\rho_2 = e^{-K}$.  The action for $K$ has
the standard free field form.  As we are in flat $D$ dimensional space
with no other background field fluctuations, the Hamiltonian is the
standard one and the energy of the field configuration (\ref{myrho})
is
\begin{equation}
E = \frac{1}{32 \pi G_N^{D}} \int d^{D-1}x \left( (\del_t K)^2 +
  (\del_{x_i} K)^2 \right) \propto \frac{1}{G_N^D A_0} (\delta R)^2 \ d^{D-3} .
\end{equation}
The Newton constant in $D+2$ dimensions is $G_N=G_N^D A_0$, and thus
we have
\begin{equation}
E_{\textup{\scriptsize{packet}}} \sim \frac{(\delta R)^2 \ d^{D-3}}{G_N}
\hsp{.1}. \label{Denergy} 
\end{equation}
Taking the energy of each dilaton to be on the order of $1/d$, the
number of dilatons comprising the wavepacket is
\begin{equation}
\textup{\# dilatons}\sim d \times
E_{\textup{\scriptsize{packet}}} \sim \frac{(\delta R)^2 \ d^{D-2}}{G_N}
\hsp{.1} . \label{Dnumber} 
\end{equation}
When $D=2$ we note that the number of dilatons in the packet is
independent of the size of the variation $d$.  This follows from the
fact that for a one dimensional packet we can always boost the size of
the variation $d$ to any arbitrary value by, for example, giving the
brane probes an initial momentum.

Using Einstein's equation, we can model the wavepacket (see
Fig.~\ref{FigVariation}) as a gravitational wave corresponding to a
Gaussian profile in the spacetime metric.  For example when $D=2$ with
noncompact coordinates $x$ and $t$, the metric along the compact
directions (the metric is $\eta_{\mu\nu}$ along all $D$ noncompact
directions) is
\begin{equation} g_{mn}= \left[
  \begin{array}{cc}
    A + h e^{-\frac{(x-t)^2}{d^2}} & 0 \\
    0 & A + h e^{-\frac{(x-t)^2}{d^2}}
                \end{array}  \right] . \label{Tmetric}
\end{equation}
The background metric on the torus is $A\delta_{m,n}$ and the
amplitude of the wavepacket variation is parametrized by $h$.
Treating, as usual, the Einstein tensor of this gravitational wave as
an effective stress-energy tensor and then integrating over all space
we learn that
\begin{equation}
E_{\textup{\scriptsize{packet}}}=\frac{\sqrt{\pi}h^2}{2^{5/2}G_N A d }
\end{equation}
when D=2.  For general D, we find that the energy of the wavepacket is
\begin{equation}
E_{\textup{\scriptsize{packet}}} \sim \frac{h^2}{G_N A}
d^{D-3} \sim \frac{(\delta R)^2 \ d^{D-3}}{G_N} \hsp{.1},
\end{equation}
which agrees with Eq.~(\ref{Denergy}).

\begin{figure}[htb]
\begin{center}
\epsfxsize=6in\leavevmode\epsfbox{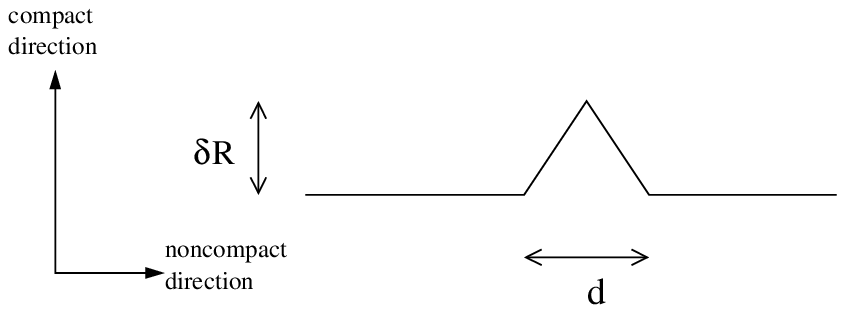}
\epsfxsize=6in\leavevmode\epsfbox{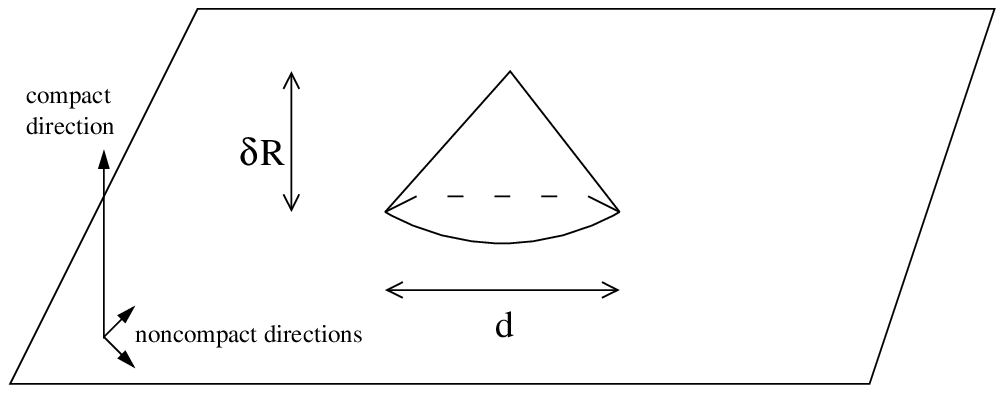}
\caption{Space as a stretched string or sheet when D=2 or 3.} \label{stretch}
\end{center}
\end{figure}

A simpler and more intuitive derivation of the results in
Eqs.~(\ref{Denergy}) and (\ref{Dnumber}) can be obtained by
considering spacetime to be a $D$-dimensional sheet with constant
tension $1/G_N$ embedded in $\R^{1,D}$.  When spacetime is Minkowskian, the flat sheet 
describes the unperturbed vacuum.  We can deform this sheet by fixing
a $D-2$ sphere in space of diameter $d$ and pulling the center of this sphere
away a distance $\delta R$ in the transverse
direction.  This locally stretches the sheet into a cone over
$S^{D-2}$ (see Fig.~\ref{stretch}).

In the case $D=2$ this sheet is just a string and the energy of this
deformation is just $1/G_N$ times how far we have stretched the
string.  When $\delta R<<d$ this is
\begin{equation}
E_{\textup{\scriptsize{packet}}} =\frac{1}{G_N}( \sqrt{(d/2)^2 - (\delta R)^2} -
d/2)\sim\frac{(\delta R)^2}{G_Nd}
\end{equation}
in agreement with Eq.~(\ref{Denergy}).  A similar calculation can be
carried out for arbitrary dimension $D$ and clearly each extra
noncompact dimension contributes one more power of $d$ because we
integrate tension over a cone of one higher dimension.

\subsection{Uncertainty Principle}

Having discussed the wavepacket we are now ready to return to the
scattering process.  Recall that we have placed just enough branes in
the path of this wavepacket so that on the order of one dilaton is
scattered.  Thus, setting the number of dilatons in the packet to be
of order one in Eq.~(\ref{Dnumber}) gives a lower bound on how small
of a $\delta R$ we can measure, or equivalently how small $\delta R$
can be for a packet of width $d$:
\begin{equation} 
(\delta R)^2 \ d^{D-2} \geq G_N \hsp{.1} .   \label{prince}
\end{equation}
We interpret this result as a space-space uncertainty.  For example,
for a single dilaton, $d$ is just the uncertainty in its position and
so Eq.~(\ref{prince}) assumes a familiar form.  In the dimensionally
reduced theory, the variation in moduli appears to be a packet of
energy moving at the speed of light in the non-compact directions with
some momentum $p\sim\delta R/d$ and so Eq.~(\ref{prince}) resembles
the Heisenberg uncertainty relation.  As such space-space uncertainty
principles are expected in all theories of quantum gravity, we will
need to learn to distinguish this effect from our conjecture in the case of more general Calabi-Yau
compactifications. 

Notice that this uncertainty principle tells us that any local
measurement of a period in some region $U \subset \R^{1,D-1}$ will
affect its value in that region.  However, it does not imply that this
measurement affects the values of other periods in $U$ and thus cannot
imply our conjecture.

\newsection{Possible Evidence on the Four-Torus} \label{t4}

The allowed scattering channels described in Sec.~\ref{SecCalc}
provide preliminary evidence for our conjecture on the four-torus.
Consider now a ten dimensional Minkowskian spacetime with global
coordinates $x_\mu$, compactified on $T^4$, which extends along
dimensions numbered $6$ through $9$.  Among the moduli of the
four-torus are the complex structure moduli of the two subtori that
span the 6-7 and 8-9 directions.  Imagine that a wavepacket
consisting of a variation of the 8-9 torus complex structure (that
is, gravitons polarized along the 8 and 9 directions) is propagating
through space, and consider measurements of the two complex structures
using 1-branes wrapped around a cycle of the subtorus to be measured.
Notice that when a 1-brane wraps a cycle of one of the tori, the other
torus sees this 1-brane as a 0-brane and thus its complex structure is
not effected by this measurement.

More precisely, choose coordinates such that the graviton scattered by
the first measurement has momentum
\begin{equation}
\hat{p}_2=\hat{x}_1.  
\end{equation}
and so the incoming wavepacket has initial momentum and polarization
\begin{equation}
\hat{p}_1=\cos(\theta)\hat{x}_1+\sin(\theta)\hat{x}_2\sp
\epsilon_1= \left( \begin{array}{ll}  i & \hsp{.1} 1 \\ 1 & -i
  \end{array}\right)
\end{equation}
where the polarization lies along the 8 and 9 directions.  To exhibit our conjecture we will
measure the complex structure variation of the subtori in both orders
and compare the results.  These measurements will be performed in a finite region $U$ and so the space-space uncertainty principle (\ref{prince}) will apply in both cases.

Let us first measure the complex structure modulus on the 6-7 torus
using 1-brane probes wrapped around the 6 direction.  Using our
results from the two-torus, we know that a complex structure variation
with these momenta can only scatter to its original polarization, or
to a dilaton or graviton with at least one polarization along the
$\hat{x}_2$ direction.  These amplitudes are invariant under time
reversal, and so only a 6-7 graviton or a dilaton or graviton with a
polarization component along the $\hat{x}_2$ direction can scatter
with our brane to produce 6-7 gravitons, thereby creating a variation
in the complex structure of this torus.  However, no such strings are
present and so this measurement does not affect the complex structure
at string tree level (although the space-space uncertainty limits the accuracy to which this measurement is possible).

\begin{figure}[htb]
\begin{center}
\epsfxsize=6in\leavevmode\epsfbox{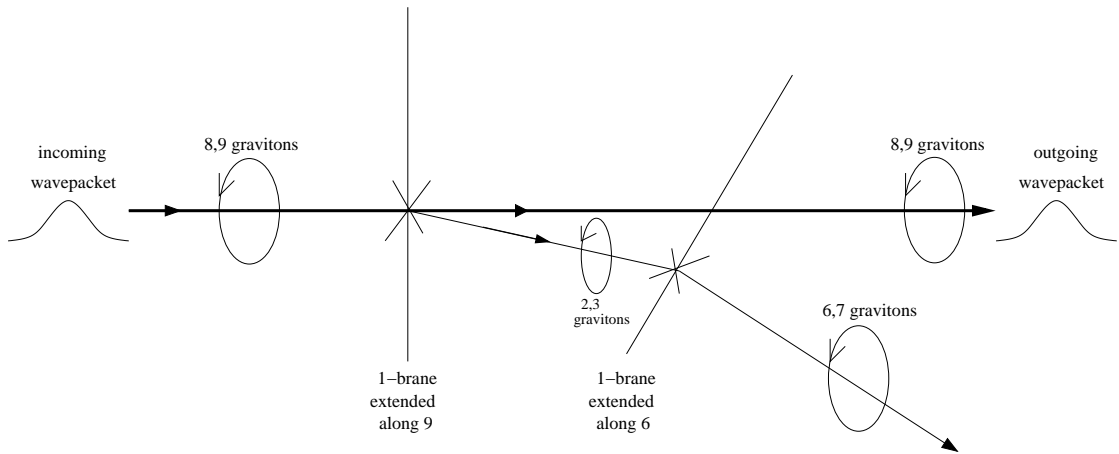}
\caption{An interaction between two complex structure moduli in T$^4$.} \label{T4}
\end{center}
\end{figure}

Alternately, let us measure the complex structure of the 8-9 torus
first (see Fig.~\ref{T4}), using 1-brane probes wrapped around the 9
direction.  While most of our 8-9 gravitons will pass by the brane
without interacting, those that scatter may change polarization to
become, for example, 2-3 gravitons.  Later, when we use 1-brane probes
wrapped around the 6 direction to measure the complex structure of the
other torus, this brane will scatter some of the 2-3 gravitons created
in the first measurement into 6-7 gravitons.  Thus, as the complex
structure of the 6-7 torus is measured, it will change.  This implies
that the result of this measurement and all subsequent measurements of
the 6-7 complex structure will be altered\footnote{In fact, if we
alternate wrappings on the brane probes we believe that the complex
structure variations of the two subtori will approach each other.  In
the case of the six-torus, this may be seen, perhaps coincidentally,
as a flow towards the attractor point identified in
Ref.~\cite{FerraraAttract3}.} if we first measure the 8-9 complex
structure, which is exactly our conjecture.  Like the space-space uncertainty principle, this effect requires the region $U$ to be finite.  However, unlike the space-space uncertainty, this effect grows with the number of 8-9 gravitons.

This measurement process can be reinterpreted as follows.  Two
necessarily non-parallel branes exchange closed strings, altering the
moduli of the branes.  The necessity of making them non-parallel, or in
the general Calabi-Yau case of wrapping the branes around different
cycles, is likely to destroy the supersymmetry.  Thus one may expect
both nontrivial interactions between the branes and that the moduli
flow as the constraints of supersymmetry are removed.

Several issues still need to be addressed before this evidence should
be taken seriously.  For example, the branes are necessarily close to
each other, and so one must check that the perturbative treatment is
valid.  Also, there may be a better way to perform this experiment,
perhaps a way to filter out the 2-3 gravitons between measurements.
If all goes well, these and other issues will be resolved in a sequel.

\noindent
{\bf Acknowledgements} 

\noindent
We express eternal gratitude to the people who were willing to speak
with us, such as N. Arkani-Hamed, K. Bardak\c{c}i, T. Bartles,
C. Callan, B. Greene, S. Kachru, R. Kallosh, M. Kleban, J. McGreevy,
S. Shenker, B. Zumino and Y. Zunger.

John and Jarah would like to thank the good people of the
\textit{Venetian Resort and Casino} for their generosity and
hospitality throughout the year.

\bibliographystyle{ieeetr}
\bibliography{escape}


\end{document}